\documentclass[pra, superscriptaddress, twocolumn, showpacs]{revtex4}
\usepackage{amssymb}
\usepackage{amsfonts}
\usepackage{epsfig}
\usepackage{amsmath}
\usepackage{graphicx}

\def\be{ \begin{equation} }
\def\ee{ \end{equation} }
\def\bea{ \begin{eqnarray} }
\def\eea{ \end{eqnarray} }
\def\bse{ \begin{subequations} }
\def\ese{ \end{subequations} }

\def\L{\Lambda}

\usepackage{calc}
\usepackage{bm}
\usepackage{color}
\usepackage{epsfig}
\usepackage{ifthen}

\begin{document}

\title{Broadband adiabatic conversion of light polarization}

\author{A. A. Rangelov}
\affiliation{Department of Physics, Sofia University, James
Bourchier 5 blvd, 1164 Sofia, Bulgaria}

\author{U. Gaubatz}
\affiliation{Nokia Siemens Networks GmbH $\And $ Co. KG,
St.-Martin-Strasse 76, 81541 Munich, Germany}

\author{N. V. Vitanov}
\affiliation{Department of Physics, Sofia University, James Bourchier 5 blvd, 1164 Sofia, Bulgaria}

\begin{abstract}
A broadband technique for robust adiabatic rotation and conversion
of light polarization is proposed. It uses the analogy between the
equation describing the polarization state of light propagating
through an optically anisotropic medium and the Schr\"{o}dinger
equation describing coherent laser excitation of a three-state
atom. The proposed technique is analogous to the stimulated Raman
adiabatic passage (STIRAP) technique in quantum optics;
 it is applicable to a wide range of frequencies and it is robust to variations in the propagation length and the rotary power.
\end{abstract}


%42.81.Gs Birefringence, polarization
%32.80.Xx Level crossing and optical pumping
%42.25.Ja Polarization
%42.25.Lc Birefringence

\maketitle

%%%%%%%%%%%%%%%%%%%%%%%%%%%%%%%%%%%%%%%%%%%%%%%%%%%%%%%%%%%%%%%%%%%%%%%%%%%%%%%%%%%%%%%%%%%%%%%%%%%%%%%%%%%%%%%%%%%%%%%%%%%%%%%
\section{Introduction}
%%%%%%%%%%%%%%%%%%%%%%%%%%%%%%%%%%%%%%%%%%%%%%%%%%%%%%%%%%%%%%%%%%%%%%%%%%%%%%%%%%%%%%%%%%%%%%%%%%%%%%%%%%%%%%%%%%%%%%%%%%%%%%%

A simple way to describe the polarization state of light, which has been known for many years in optics,
 is by the Stokes vector, which is depicted as a point on the so-called Poincar\'{e} sphere \cite{Born,Kubo80,Kubo81,Kubo83,Rothmayer09}.
The Stokes vector, for instance, is a particularly convenient tool
to describe the change of the polarization state of light
 transmitted through anisotropic optical media \cite{Kubo80,Kubo81,Kubo83}.

The equation of motion for the Stokes vector in a medium with zero polarization dependent loss (PDL) has a torque form \cite{Sala,Gregori,Tratnik}.
This fact has been used recently to draw analogies between the motion of the Stokes vector and a spin-$\frac12$ particle in
nuclear magnetic resonance and an optically driven two-state atom in quantum
optics, both described by the Schr\"{o}dinger equation \cite{Kuratsuji98,Zapasskii,Seto05,Kuratsuji07}.

Here we propose a technique for controlled robust conversion of the polarization of light
 transmitted through optically anisotropic media with no PDL.
The technique is analogous to stimulated Raman adiabatic passage (STIRAP) in quantum optics \cite{Gau90,Ber98,Vit01b}
and hence enjoys the same advantages as STIRAP in terms of efficiency and robustness.

For any traditional polarization devices the rotary power (the phase delay between the \emph{fast} and \emph{slow} eigenpolarizations)
 scales in proportion to the frequency of the light and thus such devices are frequency dependent:
 a half-wave plate is working for exactly one single frequency.
On the contrary, the adiabatic polarization conversion proposed here is \emph{frequency independent}:
 any input polarization will be transformed to the same output polarization state regardless of the wavelength.
It acts intrinsically as a broadband device limited only by the absorptive characteristics of the device instead of its birefringence bandwidth.
Moreover, the proposed technique is robust against variations in the \emph{length} of the device,
 in the same fashion as quantum-optical STIRAP is robust against variations in the pulse duration.

%%%%%%%%%%%%%%%%%%%%%%%%%%%%%%%%%%%%%%%%%%%%%%%%%%%%%%%%%%%%%%%%%%%%%%%%%%%%%%%%%%%%%%%%%%%%%%%%%%%%%%%%%%%%%%%%%%%%%%%%%%%%%%%
\section{Stokes polarization vector}
%%%%%%%%%%%%%%%%%%%%%%%%%%%%%%%%%%%%%%%%%%%%%%%%%%%%%%%%%%%%%%%%%%%%%%%%%%%%%%%%%%%%%%%%%%%%%%%%%%%%%%%%%%%%%%%%%%%%%%%%%%%%%%%

\begin{figure}[tb]
\centerline{\epsfig{width=\columnwidth,file=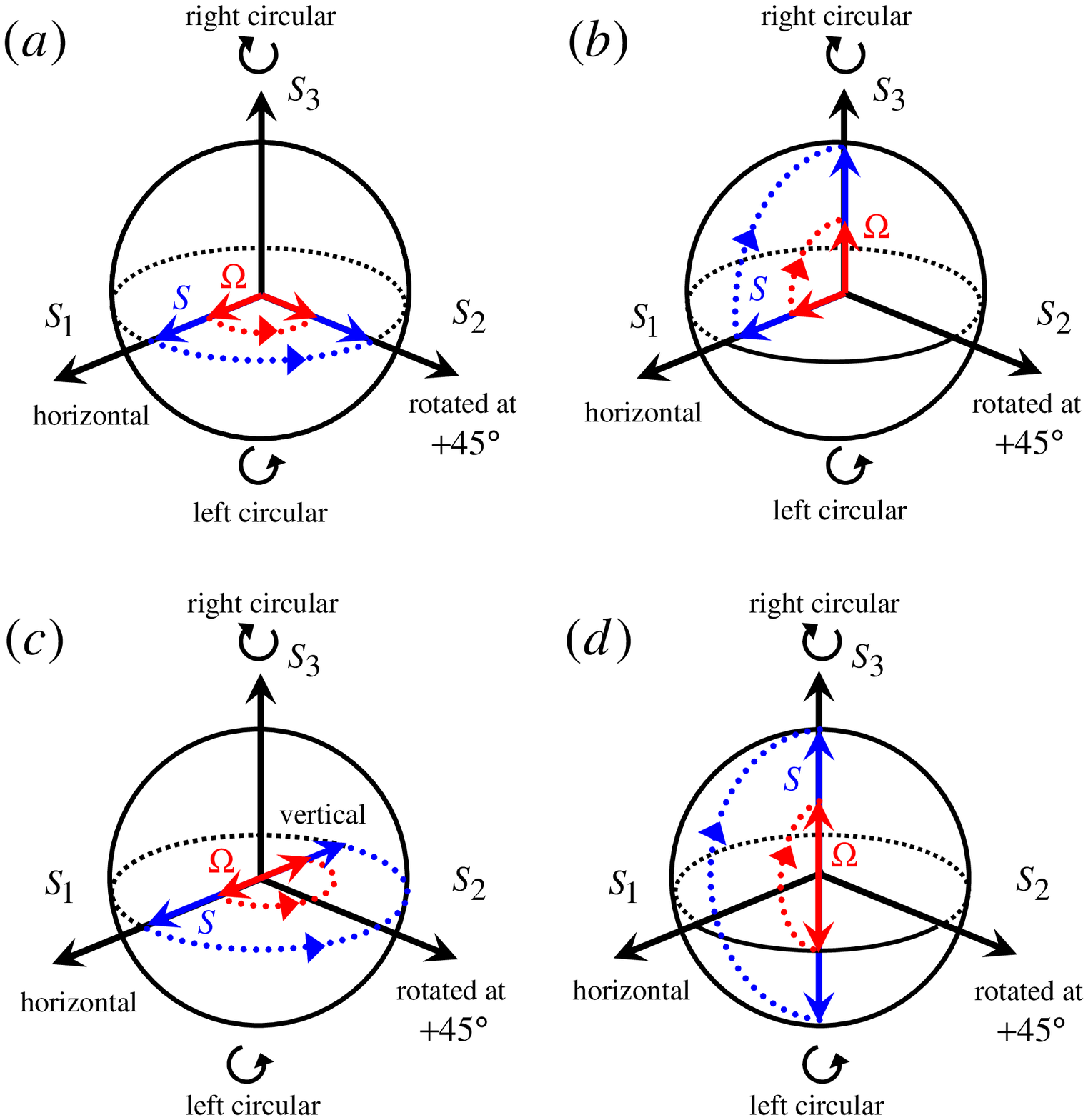}}
\caption{(Color online) Poincar\'{e} sphere representation of polarization states.
The Stokes polarization vector $\mathbf{S}(z)$ follows adiabatically the birefringence vector $\mathbf{\Omega }(z)$.
(a) Linear polarization is rotated by $45^{\circ}$ (corresponding to $90^{\circ}$ rotation of the Stokes vector);
(b) linear polarization is transferred into right circular polarization;
(c) horizontal linear polarization is transferred into vertical linear polarization;
(d) left circular polarization is transferred into right circular polarization.
} \label{Poinacare sphere}
\end{figure}

We first consider a plane electromagnetic wave traveling through a dielectric medium in the $z$ direction.
The medium is anisotropic and with no PDL, therefore the polarization evolution is given with the following torque equation
 for the Stokes vector \cite{Kubo80,Kubo81,Kubo83,Sala,Gregori,Tratnik,Kuratsuji98,Seto05,Kuratsuji07}:
\begin{equation}
\frac{\,\text{d}}{\,\text{d}z}\mathbf{S}(z)=\mathbf{\Omega }(z)\times \mathbf{S}(z),  \label{Stokes equation}
\end{equation}
 where $z$ is the distance along the propagation direction, and $\mathbf{S}(z)=[S_1(z),S_2(z),S_3(z)]$
 is the Stokes polarization vector shown in Fig. \ref{Poinacare sphere}.
Every Stokes polarization vector corresponds to a point on the Poincar\'{e} sphere and vice versa.
The right circular polarization is represented by the north pole, the left circular polarization by the south pole,
 the linear polarizations by points in the equatorial plane,
  and the elliptical polarization by the points between the poles and the equatorial plane.
$\mathbf{\Omega }(z)=[\Omega_1(z),\Omega_2(z),\Omega_3(z)]$ is the birefringence vector of the medium:
 the direction of $\mathbf{\Omega }(z)$ is given by the \emph{slow} eigenpolarization
 and its length $|\mathbf{\Omega}(z)|$ corresponds to the rotary power.

%%%%%%%%%%%%%%%%%%%%%%%%%%%%%%%%%%%%%%%%%%%%%%%%%%%%%%%%%%%%%%%%%%%%%%%%%%%%%%%%%%%%%%%%%%%%%%%%%%%%%%%%%%%%%%%%%%%%%%%%%%%%%%%
\section{Schr\"{o}dinger equation for a  three-state quantum system}
%%%%%%%%%%%%%%%%%%%%%%%%%%%%%%%%%%%%%%%%%%%%%%%%%%%%%%%%%%%%%%%%%%%%%%%%%%%%%%%%%%%%%%%%%%%%%%%%%%%%%%%%%%%%%%%%%%%%%%%%%%%%%%%

When one of the components of the vector $\mathbf{\Omega }(z)$ is zero, then Eq. \eqref{Stokes equation} is mathematically equivalent
 to the Schr\"{o}dinger equation for a coherently driven three-state quantum $\Lambda$ system
 on exact resonances with the carrier frequencies of the external fields.
This is readily seen by examining the time-dependent Schr\"{o}dinger equation, % for a resonant three-state $\Lambda $ system.
 which in the rotating-wave approximation (RWA) reads %(in units $\hbar =1$)
\begin{equation}
\text{i}\hbar \frac{\text{d}}{\text{d}t}\mathbf{c}(t)=\mathbf{H}(t)\mathbf{c}(t).  \label{Shrodinger equation}
\end{equation}
Here $\mathbf{c}(t)$ is a column vector of the probability amplitudes $c_{n}(t)$ ($n=1,2,3$) of the three states $\psi_1$, $\psi_2$ and $\psi_3$,
 and $\mathbf{H}(t)$ is a $3\times 3$ Hamiltonian matrix \cite{Gau90,Ber98,Vit01b},
\begin{equation}
\mathbf{H}(t)= \frac\hbar 2 \left[
\begin{array}{ccc}
0 & \Omega_1(t) & 0 \\
\Omega_1(t) & 0 & \Omega_2(t) \\
0 & \Omega_2(t) & 0%
\end{array}%
\right] .  \label{eqn-W}
\end{equation}%
We have here assumed both two-photon and single-photon resonances.
The two slowly varying Rabi frequencies $\Omega_1(t)$ and $\Omega_2(t)$ parameterize the strengths of each of the two fields;
 they are proportional to the dipole transition moments $\mathbf{d}_{ij}$ and to the electric-field amplitudes $\mathbf{E}_{k}(t)$:
 $\Omega_1(t) =-\mathbf{d}_{12}\mathbf{\cdot E}_1(t)$ and $\Omega_2(t)=-\mathbf{d}_{23}\mathbf{\cdot E}_2(t)$;
 hence each of them varies as the square root of the corresponding intensity.

%***************************************************************
%***************************************************************

The quantum evolution associated with the $\Lambda $ system is easily understood with the use of adiabatic states,
 i.e. the three instantaneous eigenstates $\varphi_{k}(t)$ of the RWA Hamiltonian \eqref{eqn-W}.
The adiabatic state $\varphi_0(t)$ corresponding to a zero eigenvalue is particularly noteworthy, because it has no component of state $\psi_2$,
\begin{equation}
\varphi_0(t) = \frac{\Omega_2(t)}{\sqrt{\Omega_1^2(t)+\Omega_2^2(t)}}\psi_1
 - \frac{\Omega_1(t)}{\sqrt{\Omega_1^2(t)+\Omega_2^2(t)}}\psi_3.  \label{STIRAP-dark}
\end{equation}
This state therefore does not lead to fluorescence and is known as a \emph{dark} state \cite{Gau90,Ber98,Vit01b}.
If the motion is \emph{adiabatic} \cite{Gau90,Ber98,Vit01b}, and if the state vector $\Psi (t)$ is initially
aligned with the adiabatic state $\varphi_0(t)$, then the state vector
remains aligned with $\varphi_0(t)$ throughout the evolution.
This occurs if the pulses are ordered counterintuitively, $\Omega_2(t)$ before $\Omega_1(t)$:
 then the composition of the dark state $\varphi (t)$ will progress from initial alignment with $\psi_1$ to final alignment with $\psi_3$.
Hence in the adiabatic regime the complete population will be
transferred adiabatically from state $\psi_1$ to state $\psi_3$.
This important feature of complete transfer under full control has
made STIRAP a widespread preparation technique for experiments
relying on precise state control \cite{Ber98,Vit01b}. The
formalism can be applied to other systems, including classical
ones \cite{Suchowski08,Suchowski09,Rangelov09},
 by making use of the similarity of the respective equations to Eqs.~\eqref{Shrodinger equation} and \eqref{eqn-W}.

%%%%%%%%%%%%%%%%%%%%%%%%%%%%%%%%%%%%%%%%%%%%%%%%%%%%%%%%%%%%%%%%%%%%%%%%%%%%%%%%%%%%%%%%%%%%%%%%%%%%%%%%%%%%%%%%%%%%%%%%%%%%%%%
\section{Adiabatic conversion of light polarization}
%%%%%%%%%%%%%%%%%%%%%%%%%%%%%%%%%%%%%%%%%%%%%%%%%%%%%%%%%%%%%%%%%%%%%%%%%%%%%%%%%%%%%%%%%%%%%%%%%%%%%%%%%%%%%%%%%%%%%%%%%%%%%%%

Returning to light polarization, two special cases are particularly interesting.

\emph{\textbf{Case A:}} $\Omega_3(z)=0$.
With the redefinition of variables $S_1(t)=-ic_3(t),\quad S_2(t)=ic_1(t),\quad S_3(t)=-c_2(t)$,
 the Schr\"{o}dinger equation \eqref{Shrodinger equation} turns into the form \eqref{Stokes equation},
  if we map the time dependance into the coordinate dependance.
By using this analogy we can write down a superposition $\sigma (z)$ of the polarization components $S_1(z)$ and $S_2(z)$ of the Stokes vector,
 which corresponds to the \emph{dark} state $\varphi_0(t)$, Eq.~\eqref{STIRAP-dark}:
\begin{equation}
\sigma (z)=\frac{\Omega_1(z)S_1(z)+\Omega_2(z)S_2(z)}{|\mathbf{\Omega}(z)|}.  \label{eqn-dark1}
\end{equation}

%***************************************************************
\begin{figure}[tb]
\centerline{\epsfig{width=\columnwidth,file=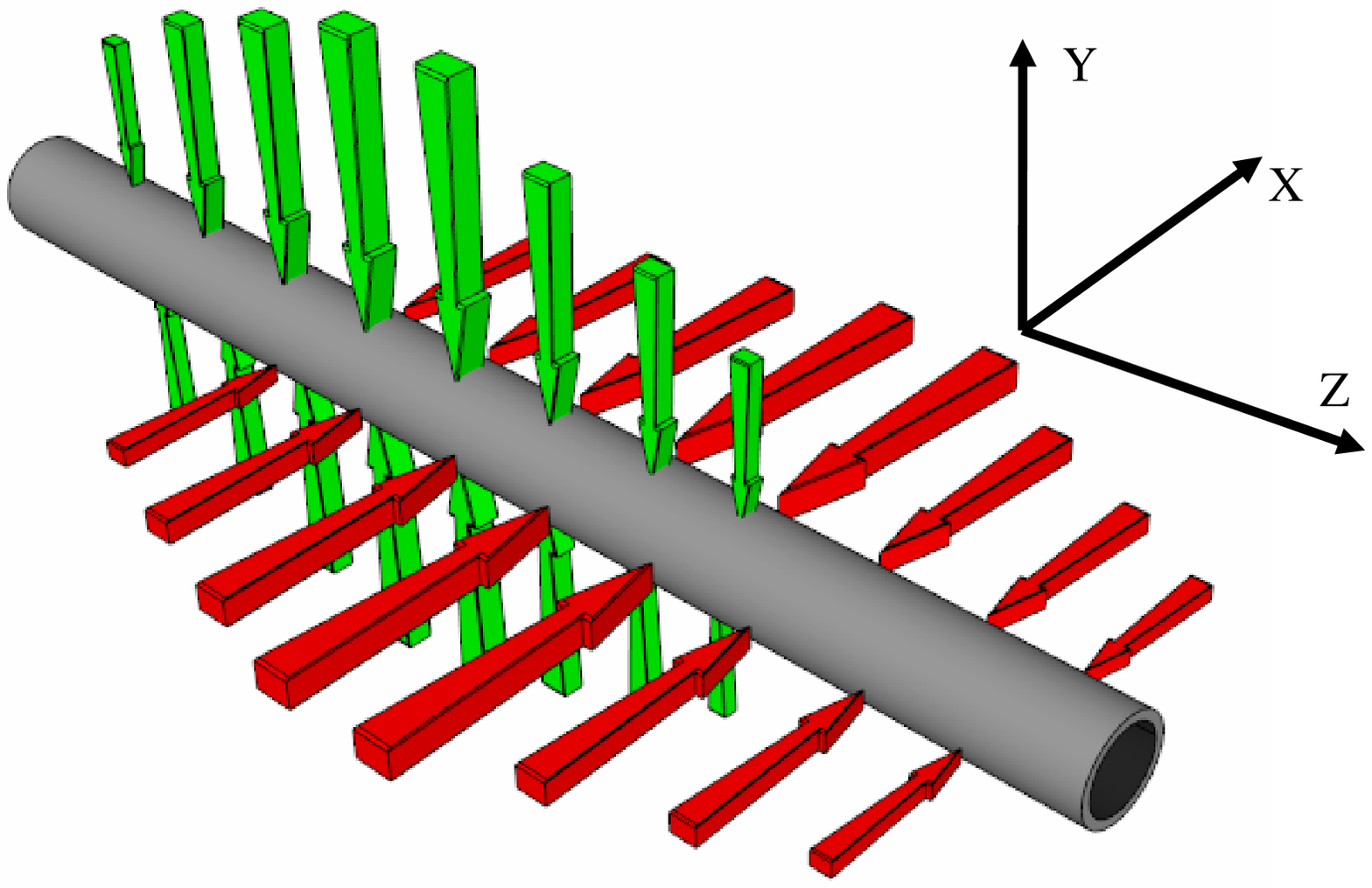}}
\caption{(Color online) Fiber optics setup with overlapping
horizontal and vertical linear birefringent sections induced by z
dependent stress (Case A). The magnitude of horizontal and
vertical stress is represented by initially increasing then
decreasing, in length, red and green arrows. } \label{Torsion}
\end{figure}
%***************************************************************
When $\Omega_1(z)$ precedes $\Omega_2(z)$ the polarization superposition \eqref{eqn-dark1} has the following asymptotic values
\begin{equation}
S_1(z_{i})\overset{z_{i}\leftarrow z}{\longleftarrow }\sigma (z)\overset{z\rightarrow z_{f}}{\longrightarrow }S_2(z_{f}).
\end{equation}
Thus if initially the light is linearly polarized in horizontal direction, $\mathbf{S}(z_{i})=(1,0,0)$, we end up with a linearly polarized light
rotated by $45^{\circ }$, $\mathbf{S}(z_{f})=(0,1,0)$ (see Fig.~\ref{Poinacare sphere} (a)).
The process is reversible; hence if we start with a linear polarization rotated by $45^{\circ }$ and apply a reverse field
order ($\Omega_2(z)$ preceding $\Omega_1(z)$) this will lead to
reversal of the direction of motion and we will end up with a linear polarization in a horizontal direction.

Following quantum-optical STIRAP, we can write down the condition for adiabatic evolution as \cite{Ber98,Vit01b}
\begin{equation}
\int_0^{L}|\mathbf{\Omega }(z)| \,dz\gg 1.
\end{equation}
For example, for adiabatic evolution it suffices to have $\int_0^{L}|\mathbf{\Omega}(z)| \,dz\gtrsim 6\pi $.
Here $L$ is the thickness of the medium and the length of the birefringence vector is given by $| \mathbf{\Omega }(z)| =2\pi \Delta n/\lambda $,
 where $\lambda $ is the wavelength of the light, $\Delta n$ is the difference between the refractive indices for ordinary and extraordinary rays.
Then the adiabatic condition reads
\begin{equation}
L\Delta n \gtrsim 3\lambda.  \label{adiabatic condition}
\end{equation}%
The last condition shows that the process is robust against variation in the parameters,
 for example, the wavelength $\lambda$ and the propagation length $L$.
This condition is readily fulfilled, by orders of magnitude, in many birefringent materials.

\emph{\textbf{Case B:}} $\Omega_2(z)=0$. Following a similar
argumentation as for \emph{\textbf{Case A}} and interchanging the
Stokes vector components $S_1(z)$ and $S_3(z)$
 we can write down a ``dark'' superposition $\sigma (z)$ of the polarization components $S_1(z)$ and $S_3(z)$ of the Stokes vector,
\begin{equation}
\sigma (z)=\frac{\Omega_1(z)S_1(z)+\Omega_3(z)S_3(z)}{\left\vert \mathbf{\Omega }(z)\right\vert }.  \label{eqn-dark2}
\end{equation}
If initially the light is linearly polarized, $\mathbf{S}(z_{i})=(1,0,0)$, then by arranging $\Omega_1(z)$ to precede $\Omega_3(z)$,
 we end up with a right circular polarization, $\mathbf{S}(z_{f})=(0,0,1)$,
 as depicted by the north pole on the Poincar\'{e} sphere in Fig.~\ref{Poinacare sphere} (b),
 because the polarization superposition \eqref{eqn-dark2} has the asymptotic values
\begin{equation}
S_1(z_{i})\overset{z_{i}\leftarrow z}{\longleftarrow }\sigma (z)\overset{z\rightarrow z_{f}}{\longrightarrow} S_3(z_{f}).
\end{equation}
The process is again reversible: starting with a right circular polarization, and applying $\Omega_3(z)$ before $\Omega_1(z)$
 we end up with a linear polarization.

We note that a numerical prediction of broadband conversion from
circular polarized light
 into linearly polarized light for the wavelength range 434 nm to 760.8 nm for crystalline quartz was made by Darsht \emph{et al.} \cite{Darsht}.
Here this effect emerges as a special case of our general
adiabatic frame of STIRAP analogy.

%***************************************************************
\begin{figure}[tb]
\centerline{\epsfig{width=\columnwidth,file=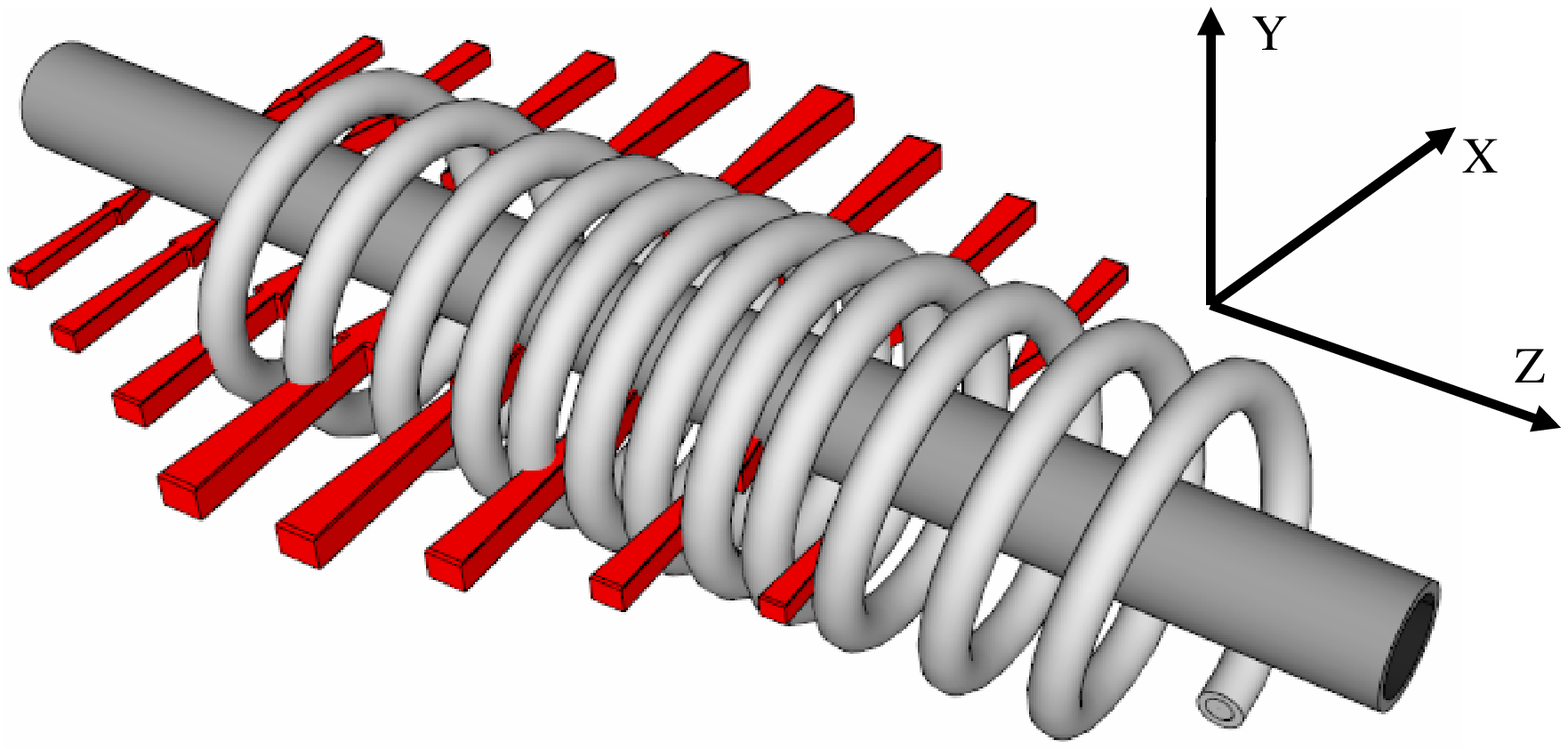}}
\caption{(Color online) Similar to Fig.~\ref{Torsion} except that
the stress induced linear birefringence has been replaced by an
circular birefringent section generated through the Faraday effect
and a magnetic field along z direction (Case B). The increasing
and decreasing number of turns in the spool indicate changes in
magnetic field.} \label{Faraday Rotation}
\end{figure}
%***************************************************************

The adiabatic polarization conversion could be demonstrated with a single-mode fiber,
 which exhibits both stress-induced linear birefringence and circular birefringence
  (either by the Faraday effect or by a torsion of the fiber) \cite{Born}.
Possible implementations are depicted in Figs.~\ref{Torsion} (for Case A) and \ref{Faraday Rotation} (for Case B).

The technique proposed here is not limited to linear-linear or circular-linear conversions,
 but it is also applicable for \emph{arbitrary} transformations of light polarization.
For example the conversion between right circular and left circular polarization
 is analogous to adiabatic passage via a level crossing \cite{Vit01b}.
To this end, we first  start up with $\Omega_3(z)$, then we
activate $\Omega_1(z)$, then let $\Omega_3(z)$ change sign
  while $\Omega_1(z)$ is having its maximal value, and then gradually make $\Omega_3(z)$ to fade
  away.

We can also change the polarization from linear to elliptical if
 we first begin with $\Omega_1(z)$, then we activate
$\Omega_3(z)$,
 and then let $\Omega_1(z)$ and $\Omega_3(z)$ \emph{simultaneously} fade away  [cf. Eq. \eqref{eqn-dark2}];
 this sequence is known in quantum optics as fractional STIRAP \cite{Half-STIRAP}.

%%%%%%%%%%%%%%%%%%%%%%%%%%%%%%%%%%%%%%%%%%%%%%%%%%%%%%%%%%%%%%%%%%%%%%%%%%%%%%%%%%%%%%%%%%%%%%%%%%%%%%%%%%%%%%%%%%%%%%%%%%%%%%%
\section{Exact solution}
%%%%%%%%%%%%%%%%%%%%%%%%%%%%%%%%%%%%%%%%%%%%%%%%%%%%%%%%%%%%%%%%%%%%%%%%%%%%%%%%%%%%%%%%%%%%%%%%%%%%%%%%%%%%%%%%%%%%%%%%%%%%%%%
%***************************************************************
\begin{figure}[h]
\centerline{\epsfig{width=\columnwidth,file=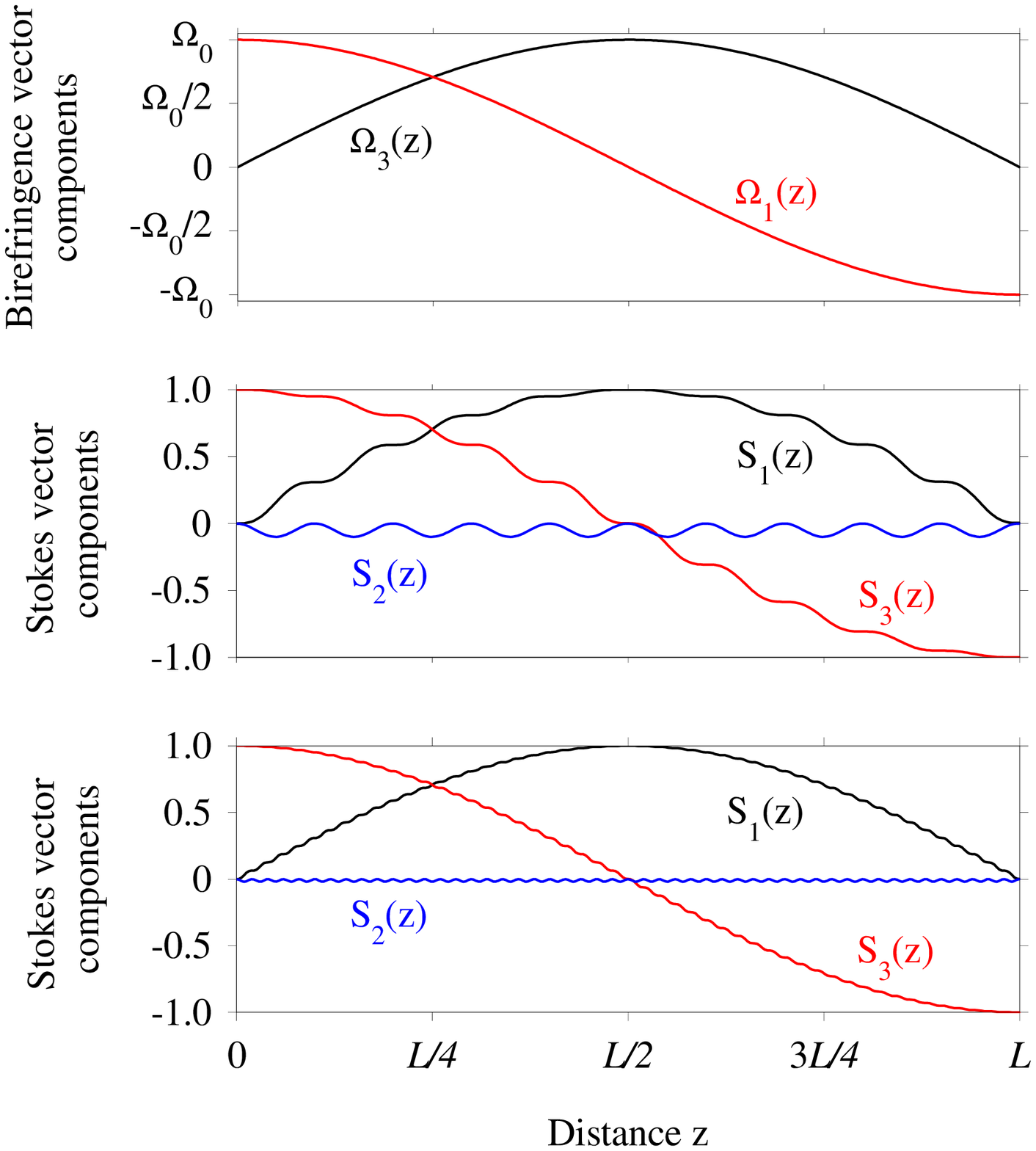}}
\caption{(Color online) Top frame: birefringence vector components
as a function of the propagation length when the components are
given as in Eq. (\ref{trigonometric functions}).  Middle frame:
the evolution of the Stokes polarization vector components for
$\Omega_0=20/L$, adiabatic evolution start to work. Bottom frame:
the evolution of the Stokes polarization vector components for
$\Omega_0=100/L$ with better adiabatic evolution, smaller ripples
on the curves, compare with the middle frame.} \label{evolution}
\end{figure}
%***************************************************************
As an example of polarization conversion we present here an exact
analytic solution to the Stokes polarization equation
\eqref{Stokes equation} for the slowly varying birefringence
components given by trigonometric functions:
\begin{subequations}
\begin{align}
\Omega_1(z) &= \Omega_0 \sin (z\pi/L), \\
\Omega_3(z) &= \Omega_0 \cos (z\pi/L).
\end{align}
\label{trigonometric functions}
\end{subequations}
We assume that initially the Stokes vector is $\mathbf{S}(z=0)=(0,0,1)$, i.e. the polarization is right circular.
Then the solution for the Stokes vector components as a function of $z$ reads
\begin{subequations}
\begin{align}
S_1(z) &= \frac{\Omega_0^2 L^2 + \cos \big( \pi z \sqrt{\L^2+\Omega_0^2}\big) }{1+\Omega_0^2 L^2}\sin \left( \frac{\pi z}{L}\right) \notag \\
&-\frac{\sin \big( \pi z \sqrt{\L^2+\Omega_0^2}\big) }{\sqrt{1+\Omega_0^2 L^2}}\cos \left( \frac{\pi z}{L}\right) , \\
S_2(z) &= \frac{2\Omega_0 L}{1+\Omega_0^2 L^2} \sin^2 \left( \frac{\pi z}{2} \sqrt{\L^2+\Omega_0^2}\right), \\
S_3(z) &= \frac{\Omega_0^2 L^2 + \cos \big( \pi z \sqrt{\L^2+\Omega_0^2}\big) }{1+\Omega_0^2 L^2}\cos \left( \frac{\pi z}{L}\right) \notag \\
&+\frac{\sin \big( \pi z \sqrt{\L^2+\Omega_0^2}\big) }{\sqrt{1+\Omega_0^2 L^2}}\sin \left( \frac{\pi z}{L}\right) ,
\end{align}
\end{subequations}
where $\L=1/L$.
The adiabatic evolution takes place when $\Omega_0 L \gg 1$;

For $z=L/2$ (quarter period) we have
\begin{subequations}
\begin{align}
S_1(L/2) &= \frac{\Omega_0^2 L^2 + \cos \big( \frac\pi 2 \sqrt{1+\Omega_0^2L^2}\big) }{1+\Omega_0^2 L^2} \overset{\Omega_0 L \gg 1}{\longrightarrow} 1 , \\
S_2(L/2) &= \frac{2\Omega_0 L}{1+\Omega_0^2 L^2} \sin^2 \left( \frac{\pi}{4} \sqrt{1+\Omega_0^2L^2}\right) \overset{\Omega_0 L \gg 1}{\longrightarrow} 0, \\
S_3(L/2) &= \frac{\sin \big( \frac\pi 2 \sqrt{1+\Omega_0^2L^2}\big) }{\sqrt{1+\Omega_0^2 L^2}} \overset{\Omega_0 L \gg 1}{\longrightarrow} 0.
\end{align}
\end{subequations}
This case exemplifies the conversion between circular and linear
polarization via STIRAP-like adiabatic process, as illustrated in
Fig. \ref{evolution} (at the point $z=L/2$).

For $z=L$ (half period) we have
\begin{subequations}
\begin{align}
S_1(L) &= \frac{\sin \big( \pi z \sqrt{\L^2+\Omega_0^2}\big) }{\sqrt{1+\Omega_0^2 L^2}} \overset{\Omega_0 L \gg 1}{\longrightarrow} 0, \\
S_2(L) &= \frac{2\Omega_0 L}{1+\Omega_0^2 L^2} \sin^2 \left( \frac{\pi z}{2} \sqrt{\L^2+\Omega_0^2}\right)\overset{\Omega_0 L \gg 1}{\longrightarrow} 0, \\
S_3(L) &= -\frac{\Omega_0^2 L^2 + \cos \big( \pi z \sqrt{\L^2+\Omega_0^2}\big) }{1+\Omega_0^2 L^2} \overset{\Omega_0 L \gg 1}{\longrightarrow} -1. \\
\end{align}
\end{subequations}
This case exemplifies the conversion between right circular and left circular polarization
 via level crossing adiabatic process, which is illustrated in Fig. \ref{evolution} (at the point $z=L$).

%%%%%%%%%%%%%%%%%%%%%%%%%%%%%%%%%%%%%%%%%%%%%%%%%%%%%%%%%%%%%%%%%%%%%%%%%%%%%%%%%%%%%%%%%%%%%%%%%%%%%%%%%%%%%%%%%%%%%%%%%%%%%%%
\section{Conclusion}
%%%%%%%%%%%%%%%%%%%%%%%%%%%%%%%%%%%%%%%%%%%%%%%%%%%%%%%%%%%%%%%%%%%%%%%%%%%%%%%%%%%%%%%%%%%%%%%%%%%%%%%%%%%%%%%%%%%%%%%%%%%%%%%

In conclusion, we have shown that the powerful technique of
STIRAP, which is well-known in quantum optics,
 has an analogue in the evolution of light polarization described by the equation for the Stokes vector.
The factor that enables this analogy is the equivalence of the Schr\"{o}dinger equation
 for a resonant three-state $\Lambda $ system, to the torque equation for the Stokes vector.
The proposed technique transforms polarization with the same efficiency and robustness as STIRAP,
 therefore a polarization device based on this scheme is frequency independent and it is robust against variations
 of the propagation length, in contrast to the other well-known methods for conversion of light polarization.

This work has been supported by the European Commission projects EMALI and FASTQUAST,
 the Bulgarian NSF grants D002-90/08, DMU02-19/09 and Sofia University Grant 074/2010. We are grateful to Klaas Bergmann for useful discussions.


\begin{thebibliography}{99}

\bibitem{Born} M. Born and E. Wolf, \emph{Principles of Optics} (Pergamon, Oxford, 1975).

\bibitem{Kubo80} H. Kubo and R. Nagata,
%Stokes parameters representation of the light propagation equations in inhomogeneous anisotropic, optically active media,
 Opt. Commun. \textbf{34}, 306 (1980).

\bibitem{Kubo81} H. Kubo and R. Nagata,
%Determination of dielectric tensor fields in weakly inhomogeneous anisotropic media. II,
 J. Opt. Soc. Am. \textbf{71}, 327 (1981).

\bibitem{Kubo83} H. Kubo and R. Nagata, ``Vector representation of behavior
of polarized light in a weakly inhomogeneous medium with
birefringence and dichroism", J. Opt. Soc. Am. \textbf{73}, 1719
(1983).

\bibitem{Rothmayer09} M. Rothmayer, D. Tierney, E. Frins, W. Dultz, and H.
Schmitzer,
%Irregular spin angular momentum transfer from light to small birefringent particles,
Phys. Rev. A \textbf{80}, 043801 (2009).


\bibitem{Sala} K. L. Sala,
%Nonlinear refractive-index phenomena in isotropic media subjected to a dc electric field: Exact solutions,
 Phys. Rev. A \textbf{29}, 1944 (1984).

\bibitem{Gregori} G. Gregori and S. Wabnitz,
%New exact solutions and bifurcations in the spatial distribution of polarization in third-order nonlinear optical interactions,
 Phys. Rev. Lett. \textbf{56}, 600 (1986).

\bibitem{Tratnik} M. V. Tratnik and J. E. Sipe,
%Nonlinear polarization dynamics. I. The single-pulse equations,
 Phys. Rev. A \textbf{35}, 2975 (1987).

\bibitem{Kuratsuji98} H. Kuratsuji and S. Kakigi,
%Maxwell-Schr\"{o}dinger Equation for Polarized Light and Evolution of the Stokes Parameters,
 Phys. Rev. Lett. \textbf{80}, 1888 (1998).

\bibitem{Zapasskii} V. S. Zapasskii and G. G. Kozlov,
%Polarized light in an anisotropic medium versus spin in a magnetic field
Phys.-Usp. \textbf{42,} 817 (1999).

\bibitem{Seto05} R. Seto, H. Kuratsuji, and R. Botet,
%Resonant oscillations of the Stokes parameters in non-linear twisted birefringent media,
 Europhys. Lett. \textbf{71}, 751 (2005).

\bibitem{Kuratsuji07} H. Kuratsuji, R. Botet, and R. Seto,
%Electromagnetic Gyration,
 Prog. Theor. Phys. \textbf{117}, 195 (2007).

\bibitem{Gau90} U. Gaubatz, P. Rudecki, S. Schiemann, and K. Bergmann,
%Population transfer between molecular vibrational levels by stimulated Raman scattering with partially overlapping laser fields. A new concept and experimental results,
J. Chem. Phys. \textbf{92}, 5363 (1990).

\bibitem{Ber98} K. Bergmann, H. Theuer, and B. W. Shore,
%Coherent population transfer among quantum states of atoms and molecules,
Rev. Mod. Phys. \textbf{70}, 1003 (1998).

\bibitem{Vit01b} N. V. Vitanov, M. Fleischhauer, B. W. Shore, and K. Bergmann,
%Coherent manipulation of atoms and molecules by sequential laser pulses,
 Adv. At. Mol. Opt. Phys. \textbf{46}, 55 (2001).

\bibitem{Suchowski08} H. Suchowski, D. Oron, A. Arie, and Y. Silberberg,
%Geometrical representation of sum frequency generation and adiabatic frequency conversion,
Phys. Rev. A \textbf{78}, 063821 (2008).

\bibitem{Suchowski09} H. Suchowski, V. Prabhudesai, D. Oron, A. Arie, and Y. Silberberg,
%Robust adiabatic sum frequency conversion,
Opt. Express \textbf{17}, 12731 (2009).

\bibitem{Rangelov09} A. A. Rangelov, N. V. Vitanov, and B. W. Shore,
%Stimulated Raman adiabatic passage analogues in classical physics,
J. Phys. B \textbf{42}, 055504 (2009).

\bibitem{Darsht} M. Ya. Darsht, B. Ya. Zel'dovich, and N. D. Kundikova,
 Opt. Spektrosk. \textbf{82}, 660 (1997).

\bibitem{Half-STIRAP} N. V. Vitanov, K.-A. Suominen, and B. W. Shore,
%Creation of coherent atomic superpositions by fractional stimulated Raman adiabatic passage,
J. Phys. B \textbf{32}, 4535 (1999).

%\bibitem{Half-STIRAP} N. V. Vitanov, K.-A. Suominen, and B. W. Shore, ``Creation of coherent atomic superpositions by fractional stimulated Raman adiabatic passage", J. Phys. B \textbf{32}, 4535 (1999).
\end{thebibliography}
\end{document}